\documentclass[twocolumn,showpacs,pra,amsmath]{revtex4-1}

\usepackage{epsfig}
\usepackage{subfigure}
\usepackage{graphicx}% Include figure files
\usepackage{dcolumn}% Align table columns on decimal point
\usepackage{stmaryrd}
\usepackage{mathrsfs}
\usepackage{pifont}
\usepackage{amsthm}
\usepackage{amssymb}
\usepackage{bm}
\usepackage{latexsym}
\usepackage{hyperref}

\newcommand{\la}{\langle}
\newcommand{\ra}{\rangle}

\newcommand{\ti}{\tilde}
\newcommand{\ga}{\gamma}
\newcommand{\Ga}{\Gamma}
\newcommand{\ka}{\kappa}
\newcommand{\da}{\dagger}
\newcommand{\De}{\Delta}
\newcommand{\al}{\alpha}

\newcommand{\om}{\omega}
\newcommand{\de}{\delta}

\newcommand{\pa}{\partial}
\newcommand{\non}{\nonumber}

\def\jpa#1{{ J.\ Phys.\ A} {\bf#1}}
\def\pra#1{{ Phys.\ Rev. A\/} {\bf#1}}

\def\prl#1{{ Phys.\ Rev.\ Lett.} {\bf#1}}
\def\pr#1{{ Phys.\ Rev.} {\bf#1}}
\def\sci#1{{ Science} {\bf#1}}

\def\pla#1{{ Phys.\ Lett. A\/} {\bf#1}}
\def\rmp#1{{ Rev. \ Mod. \ Phys.} {\bf#1}}

\begin{document}

\title{Time-Local Quantum-State-Diffusion Equation for Multilevel Quantum
Dynamics}

\author{Jun Jing$^{1,2}$}\email{Jun.Jing@stevens.edu}

\author{Xinyu Zhao$^{1}$}

\author{J. Q. You$^{3}$}\email{jqyou@fudan.edu.cn}

\author{Ting Yu$^{1}$}\email{Ting.Yu@stevens.edu}

\affiliation{$^{1}$Center for Controlled Quantum Systems and Department of
Physics and Engineering Physics, Stevens Institute of Technology, Hoboken, New
Jersey 07030, USA \\ $^{2}$Department of Physics, Shanghai University, Shanghai
200444, China \\ $^{3}$Department of Physics and State Key Laboratory of
Surface Physics, Fudan University, Shanghai 200433, China}

\date{\today}

\begin{abstract}
An open quantum system with multiple levels coupled to a bosonic environment at
zero temperature is investigated systematically using the non-Markovian quantum
state diffusion (QSD) method [W. T. Strunz, L. Di\'osi and N. Gisin, Phys. Rev.
Lett. {\bf 82}, 1801 (1999)]. We have established exact time-local QSD
equations for a set of interesting multilevel open systems including the
high-spin systems, multiple transition atom models, and multilevel atom models
driven by time-dependent external fields. These exact QSD equations have paved
a way to evaluate the dynamics of the open multilevel atomic systems in the
general non-Markovian regimes without any approximation.
\end{abstract}


\maketitle

\section{Introduction}

Many quantum processes, including emission, absorption and quantum interference
\cite{SYZhu,Ficek}, typically involve multilevel atomic systems coupled to a
quantized bosonic field. For example, quantum multilevel models are used in
studies of the high-spin model \cite{Ziolo}, super-radiance, quantum phase
transition in Dicke model \cite{Dicke,Milburn}, electromagnetically induced
transparency (EIT) \cite{EITrmp}, molecular aggregate \cite{eisfield} and many
other phenomena in quantum optics \cite{Scully,C1} and quantum chemistry
\cite{Teki,Real}. More recently, multilevel atomic systems also play a crucial
role in describing coherent state transfer \cite{Spiller} and entanglement
dynamics and control \cite{Polzik,Wang,Paz,C2,C3,Yu-Eberly2010}. All of those
models concern the interactions between a multilevel atomic system and its
surrounding environment and external driving fields leading to collective and
competitive behaviors including dissipation, fluctuation, decoherence and
revival. Approximate master equations and other alternatives such as the
Langevin equations or quantum trajectories are used to describe quantum
dynamics of the multilevel atomic system when the weak-coupling and Markov
approximations are valid \cite{Breuer}.

In this paper, the multilevel atomic systems will be treated in the framework
of the non-Markovian quantum state diffusion equation
\cite{Diosi97,Diosi98,Strunz99-1,Strunz99-2,Strunz01,Jing10} without any
approximations. Unlike the methods of Markov quantum state diffusion or quantum
jump simulations \cite{Gisin,jump1,jump2,jump3,Wiseman}, where quantum
trajectories are unravelings of the density operator of the system of interest,
the non-Markovian QSD equation is derived from the first principle and is
determined uniquely by the system Hamiltonian, the coupling operator (called
the Lindblad operator) and the spectral density of the environment
\cite{Diosi98,Weiss}. It has been shown that the non-Markovian QSD equation can
be useful in the following ways: (i) it is an exact description of quantum
dynamics of the system under the influence of the environmental noise with
finite memory time \cite{Bassi,Gardiner,non-Markovian1,non-Markovian2,C4}, in
particular, it serves as a powerful tool in numerical simulations of quantum
open systems; (ii) it could also be used to derive the exact or approximate
master equations for the reduced density matrix \cite{Yu99,Strunz_Yu04,Yu04}.

Exact QSD equations have been found in several interesting models including a
two-level atom in the dissipative environment \cite{Strunz99-1}, a harmonic
oscillator in Brownian motion \cite{Strunz_Yu04}, a three-level atom in the
dissipation model \cite{Jing10}, and a two-qubit model \cite{Zhao2011}.
However, it remains unknown how exact QSD equations can be derived for general
multilevel systems coupled to a dissipative environment. The purpose of this
paper is to establish a set of exact QSD equations for a large class of
multilevel atomic models coupled to a dissipative environment with or without
external driving fields.

This paper will be organized as follows. In Sec.~\ref{qsd}, we will begin by
reviewing the basic concepts of the non-Markovian QSD approach that will be
useful for the discussions presented in the subsequent sections including
the formal QSD equation and the O-operators. In Sec.~\ref{Oop}, we will then
proceed to discuss the construction of the time-local QSD equation by
explicitly determining O-operators in various multilevel models. Afterwards, we
present the numerical results from QSD simulations to illustrate the
non-Markovian dynamics of multilevel atomic systems under environmental noises
in Sec.~\ref{discussion}. Finally we will conclude the paper in
Sec.~\ref{conclusion}.

\section{Non-Markovian QSD Equation}\label{qsd}

In the framework of system-plus-environment, the total Hamiltonian is given by
(setting $\hbar=1$):
$H_{\rm tot}=H_{\rm sys} +\sum_{\bf k}(g_{\bf k}^*La_{\bf k}^\dagger + g_{\bf
k}L^\da a_{\bf k})+\sum_{\bf k} \omega_{\bf k} a^\da_{\bf k}a_{\bf k}$, where
$H_{\rm sys}$ is the system Hamiltonian and $L$ is said to be the Lindblad
operator coupling the system to the environment. The environment is described
by a set of harmonic oscillators $a^\da_{\bf k}, a_{\bf k}$ satisfying $[a_{\bf
k}, a^\da_{\bf k'}]=\delta_{\bf k, k'}$. The non-Markovian QSD approach is
designed to recover the reduced density matrix for the open system by the
average of the pure states driven by a colored Gaussian process
$z^*_t$:
\begin{equation}\label{rho}
\rho_t=M[|\psi_t(z^*)\ra\la\psi_t(z^*)|],
\end{equation}
where $M$ stands for the statistical average over the noise $z_t^*$. The
dynamics of the stochastic unravelings $\psi_t(z^*)$ (quantum trajectories) is
governed by the non-Markovian QSD equation \cite{Diosi97,Diosi98}:
\begin{equation}\label{qsd1}
\pa_t\psi_t(z^*)=\left[-iH_{\rm sys}+Lz^*_t-
L^\da\int_0^tds\al(t,s)\frac{\de}{\de z^*_s}\right]\psi_t(z^*),
\end{equation}
where $\al(t,s)=\sum_{\bf k} |g_{\bf k}|^2 e^{-i\omega_{\bf k}(t-s)}$ is the
environmental correlation function. By construction, the Gaussian noise $z_t^*$
satisfies $M[z_t^*]=M[z_t^*z_s^*]=0$ and $M[z_tz_s^*]=\al(t,s)$. Note that the
functional derivative contained in Eq.~(\ref{qsd1}) is a time non-local term
depending on the entire evolution history from $0$ to $t$. This time non-local
term is a major obstacle in using Eq.~(\ref{qsd1}) as a numerical tool or as an
analytical approach to deriving the corresponding non-Markovian master
equation. One way to transform the formal non-Markovian QSD equation
(\ref{qsd1}) to a time-local equation is to introduce an O-operator satisfying
$\frac{\de\psi_t(z^*)}{\de z^*_s}\equiv O(t,s,z^*)\psi_t(z^*)$, then the QSD
equation could be recast into a convolutionless form:
\begin{equation}\label{qsd2}
\pa_t\psi_t(z^*)=\left[-iH_{\rm sys}+Lz^*_t
-L^\da\bar{O}(t,z^*)\right]\psi_t(z^*),
\end{equation}
where $\bar{O}(t,z^*)\equiv\int_0^tds\al(t,s)O(t,s,z^*)$ and the O-operator may
be determined from the following equation:
\begin{eqnarray}\non
\frac{\pa O(t,s,z^*)}{\pa t}&=&[-iH_{\rm sys}+Lz^*_t - L^\da\bar{O}(t,z^*), \\
\label{CC} && O(t,s,z^*)]-L^\da\frac{\de\bar{O}(t,z^*)}{\de z^*_s}.
\end{eqnarray}

Hence, the key issue in establishing the exact time-local QSD equations is to
explicitly construct the O-operator defined in (\ref{CC}). In what follows, we
shall show how to determine the O-operator explicitly for a large class of
multilevel atomic systems.

It is worthwhile to note that, from the density matrix reconstruction defined
in Eq.~(\ref{rho}) and Novikov theorem \cite{Yu99}, we may obtain an exact
master equation from Eq.~(\ref{qsd2}):
\begin{equation}
\pa_t\rho_t=[-iH_{\rm sys},\rho_t]+[L,M[\hat{P}_t\bar{O}^\da]]
+[M[\bar{O}\hat{P}_t],L^\da],
\end{equation}
where $\hat{P}_t\equiv|\psi_t(z^*)\ra\la\psi_t(z^*)|$. The master equation
takes a more concise form when the O-operator is noise-free:
$O(t,s,z^*)=O(t,s)$,
\begin{equation}\label{nfme}
\pa_t\rho_t=-i[H_{\rm sys},\rho_t]+[L,\rho_t\bar{O}^\da]+[\bar{O}\rho_t,L^\da].
\end{equation}

In particular, under the Born-Markov approximation with the delta function
$\al(t,s)=\Gamma \delta(t-s)$, then $\bar{O}(t,s,z^*)=\Gamma L/2$.
Eq.~(\ref{nfme}) recovers the Lindblad master equation:
\begin{equation}
\pa_t\rho_t=-i[H_{\rm sys},\rho_t]
+\frac{\Ga}{2}\left([L,\rho_tL^\da]+[L\rho_t,L^\da]\right).
\end{equation}

\section{Time-Local QSD equations}\label{Oop}

\subsection{High-Spin Model}

Now we consider a collective angular momentum (Dicke) model or the high-spin
model described by
\begin{equation}\label{Dick}
H_{\rm sys}=\om J_z, \quad L=J_-,
\end{equation}
where $J_z$ and $J_-$ are the angular momentum operators with spin-$l$. It can
be proved (see Appendix \ref{Oopdicke}) that the O-operator contains $l(2l+1)$
terms with the basis operators $O_j^{(k)}=|j\ra\la j+k+1|$, $j=1,\cdots,2l-k$,
for each $k=0,\cdots, 2l-1$. And the terms with $O_j^{(k)}$ contain $k$th order
noises for $k=0,\cdots, 2l-1$.

Clearly, one may construct many different sets of basis operators, but they can
always be realized by linear combinations of the operators $O_j^{(k)}$ selected
here. The basis operators may also be given by
$O_j^{(k)}=J_z^{j-1}J_-^{k+1}$. The simplest model is the spin-$1/2$ system
($l=1/2$), where the O-operator contains only one term with the basis operator
$O^{(0)}_1=\sigma_-$. Another model is spin-$3/2$. We can show the O-operator
$O(t,s,z^*)$ totally contains $6$ terms, amongst, only one term contains the
second-order noise:
\begin{eqnarray} \non
O&=&\sum_{j=1}^3f_j(t,s)O_j^{(0)}
+\sum_{j=1}^2\int_0^tp_j^{(1)}(t,s,s_1)z^*_{s_1}ds_1O_j^{(1)} \\ \label{O4l}
&&+ \int_0^t\int_0^tp_1^{(2)}(t,s,s_1,s_2)z^*_{s_1}z^*_{s_2}ds_1ds_2O_1^{(2)}.
\end{eqnarray}

In Table \ref{qutab}, we list the numbers of the basis operators for the
high-spin models from spin-$1/2$ to spin-$7/2$.

\begin{table}[h!]\centering
\begin{tabular}{|c|c|c|c|c|c|c|c|c||c|} \hline
% \backslashbox{$2l$}{$k$}
$2l \setminus k$  & $0$ & $1$ & $2$ & $3$ & $4$ & $5$ & $6$ & $7$
& $N$ \\ \hline
$1$ & $1$ & $0$ & $0$ & $0$ & $0$ & $0$ & $0$ & $0$ & $1$ \\ \hline
$2$ & $2$ & $1$ & $0$ & $0$ & $0$ & $0$ & $0$ & $0$ & $3$ \\ \hline
$3$ & $3$ & $2$ & $1$ & $0$ & $0$ & $0$ & $0$ & $0$ & $6$ \\ \hline
$4$ & $4$ & $3$ & $2$ & $1$ & $0$ & $0$ & $0$ & $0$ & $10$ \\ \hline
$5$ & $5$ & $4$ & $3$ & $2$ & $1$ & $0$ & $0$ & $0$ & $15$  \\ \hline
$6$ & $6$ & $5$ & $4$ & $3$ & $2$ & $1$ & $0$ & $0$ & $21$  \\ \hline
$7$ & $7$ & $6$ & $5$ & $4$ & $3$ & $2$ & $1$ & $0$ & $28$  \\ \hline
\end{tabular}
\caption{The number of basis operators with $k$-fold integration over noises in
the O-operator for spin-$l$ system in a dissipation model. $N$ is total number
of the terms in the O-operators.}
\label{qutab}
\end{table}

The Hamiltonian and Lindblad operator considered in Eq.~(\ref{Dick}) imply that
the energy levels of the multilevel atom are equidistant and all the exited
levels dissipate to the lower neighboring levels. However, these two constraints
can be relaxed. We start with the model Eq.~(\ref{Dick}) with the modified
Hamiltonian and coupling operator:
\begin{equation}
H_{\rm sys}=\sum_j \om_j|j\ra\la j|, \quad L=\sum_j \ka_j|j\ra\la j+1|.
\end{equation}
More specifically, we use a three-level system as an example for simplicity. It
can be verified that the O-operator can be constructed as follows (For details,
see Appendix \ref{Oopdicke} or Eq.~(6) of \cite{Jing10}):
\begin{equation}
O=f_1(t,s)J_-+f_2(t,s)J_zJ_-+\int_0^tp(t,s,s_1)z^*_{s_1}ds_1J_-^2.
\end{equation}
Consequently,
$\bar{O}(t,z^*)=F_1(t)J_-+F_2(t)J_zJ_-+\int_0^tP(t,s_1)z^*_{s_1}
ds_1J_-^2$, where $F_j(t)\equiv\int_0^t\al(t,s)f_j(t,s)ds$, $j=1,2$ and
$P(t,s_1)\equiv\int_0^t\al(t,s)p(t,s,s_1)ds$. The initial conditions for these
coefficients are $f_1(s,s)=\ka_2/\sqrt{2}$, $f_2(s,s)=(\ka_2-\ka_1)/\sqrt{2}$,
$p(s,s,s_1)=0$ and they satisfy:
\begin{eqnarray}\non
\pa_tf_1&=&i(\om_3-\om_2)f_1-\sqrt{2}[(\ka_1-\ka_2)F_1-\ka_1F_2]f_1\\
\non &-&\sqrt{2}i\ka_1P(t,s), \\ \non
\pa_tf_2&=&i(\om_3+\om_1-\om_2)f_1-\sqrt{2}i\ka_1P(t,s)
\\ \non &+&i(\om_2-\om_1)f_2+\sqrt{2}\ka_1(F_1-F_2)f_2
\\ \non &-&\sqrt{2}[(2\ka_1-\ka_2)F_1-2\ka_1F_2]f_1 , \\ \non
\pa_tp&=&i(\om_3-\om_1)p+\sqrt{2}\ka_2F_1p+\sqrt{2}\ka_1P(f_1-f_2),
\\ p(t,s,t)&=&i\frac{\ka_2-\ka_1}{\sqrt{2}}f_1(t,s)
-i\frac{\ka_2}{\sqrt{2}}f_2(t,s).
\end{eqnarray}
These equations will reduce to Eqs.~(7), (8) and (9) in \cite{Jing10} by
setting $\om_3=\om$, $\om_2=0$, $\om_1=-\om$ and $\ka_1=\ka_2=\sqrt{2}$. As
shown in the next subsection, the time-local QSD approach can be extended to
the driven atomic models with the multiple transitions between the energy
levels.

\subsection{Multiple Transition and Driven Atomic Models}

\begin{figure}[htbp] \centering
 \includegraphics[width=2.7in]{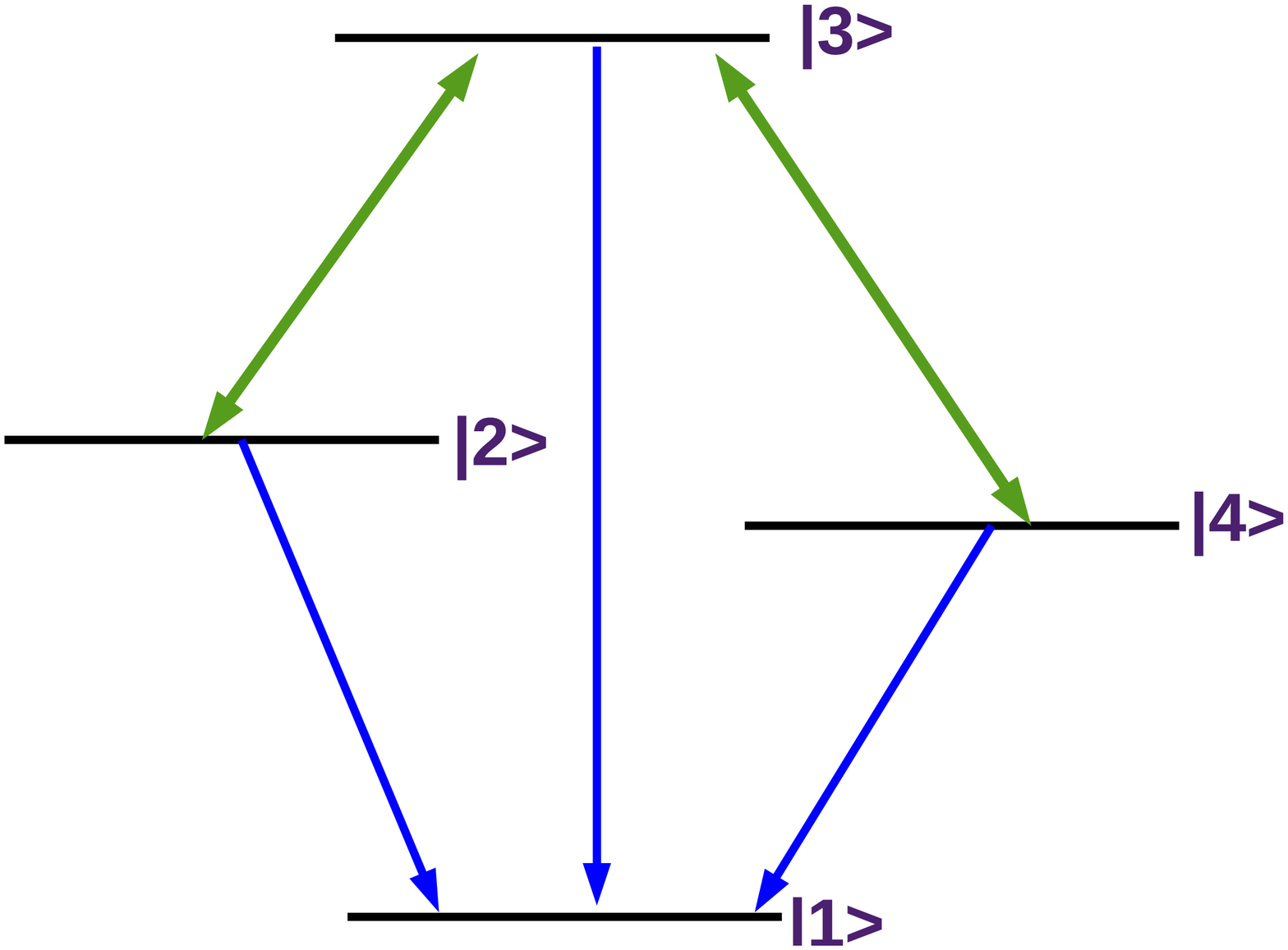}
\caption{(Color online) Schematic diagram of a general EIT model: A four-level
atomic system with two driving fields (green double arrow lines) as
$\Delta_2(t)T_{23}+h.c.$ and $\Delta_4(t)T_{34}+h.c.$. The admitted damping
channels (blue single arrow lines) are $T_{12}$, $T_{13}$ and $T_{14}$.}
\label{Diag}
\end{figure}

In this subsection, we consider a driven multi-level atom with multiple
dissipative channels. The system Hamiltonian is given by,
\begin{equation}\label{Hs}
H_{\rm sys}=\sum_{j=1}^N\omega_j|j\ra\la j|+
\sum_{k_1\neq k_2} [\Delta_{k_1k_2}(t)|k_1\ra\la k_2|+h.c.],
\end{equation}
where $\Delta_{k_1k_2}(t)$'s are the time-dependent functions and $1\leqslant
k_1, k_2 \leqslant N$. Also, we consider a general Lindblad operator, which is
given by
\begin{equation}\label{LDi}
L=\sum_{n_1\neq n_2}\kappa_{n_1n_2}|n_1\ra\la n_2|.
\end{equation}
We remark that one may always determine the exact time-local QSD equation for
the above generic Hamiltonian and the Lindblad operator. However, it can be
verified that a particularly simple noise-free O-operator exists if the
following two conditions are satisfied: (i) any operator $|k_1\ra\la k_2|$
appearing in the driving term is not present in the Lindblad operator in
Eq.~(\ref{LDi}); (ii) the cycle transition terms, for example,  $|n_1\ra\la
n_2|$, $|n_2\ra\la n_3|$ and $|n_1\ra\la n_3|$, cannot be simultaneously
contained in the Lindblad operator. As an illustration, we consider an
interesting model consisting of a four-level atom shown in Fig.~(\ref{Diag}),
the system is described by
\begin{eqnarray}\non
H_{\rm sys}&=&\sum_{j=1}^4\omega_j|j\ra\la j|+
[\Delta_{2}(t)|2\ra\la 3|
+\Delta_{4}(t)|3\ra\la 4|+h.c.], \\ \label{FDA}
L&=&\kappa_{2}|1\ra\la 2|+\kappa_{3}|1\ra\la 3|+\kappa_{4}|1\ra\la 4|,
\end{eqnarray}
where $\Delta_2(t)$ and $\Delta_4(t)$ are independent driving external fields.
Thus the O-operator can be constructed as
$O(t,s,z^*)=\sum_{j=2}^4f_j(t,s)|1\ra\la j|$ with $f_j(s,s)=\ka_j$. And the
differential equations for the coefficients are
\begin{eqnarray}\non
\pa_tf_2(t,s)&=&-i\om_{12}f_2+i\De_2^*f_3+F_2\sum_{j=2}^4\ka_j^*f_j, \\ \non
\pa_tf_3(t,s)&=&-i\om_{13}f_3+i\De_2f_2+i\De_4^*f_4+F_3\sum_{j=2}^4\ka_j^*f_j
\\ \pa_tf_4(t,s)&=&-i\om_{14}f_4+i\De_4f_3+F_4\sum_{j=2}^4\ka_j^*f_j,
\end{eqnarray}
where $\om_{1j}\equiv\om_1-\om_j$ and $F_j\equiv
F_j(t)=\int_0^tds\alpha(t,s)f_j(t,s)$, $j=2,3,4$.

When the driving terms in $H_{\rm sys}$ are omitted, the model in
Eq.~(\ref{Hs}) reduces to an $N$-level atom with multiple transition channels
\cite{Curtis} where the transition takes place between the highest level and
all the lower energy levels, and transitions between any other levels are
forbidden. Explicitly, the Hamiltonian and Lindblad operator in this case are
given by
\begin{equation}
H_{\rm sys}=\sum_{j=1}^N\om_j|j\ra\la j|,
\quad L=\sum_{j=1}^{N-1}\ka_j|j\ra\la N|,
\end{equation}
respectively. For this model, we can show that the O-operator can be explicitly
constructed as $O(t,s,z^*)=\sum_{j=1}^{N-1}f_j(t,s)|j\ra\la N|$. Hence
$\bar{O}(t,z^*)=\sum_{j=1}^{N-1}F_j(t)|j\ra\la N|$ with the initial conditions,
$f_j(s,s)=\ka_j$, $j=1,\cdots,N-1$. By the consistency condition (\ref{CC}),
one gets
\begin{equation}
\pa_tf_j(t,s)=i(\om_N-\om_j)f_j+f_j\sum_{k=1}^{N-1}\ka_k^*F_k(t).
\end{equation}

In a more general case, we may consider an $(N+M)$-level atom with an upper
energy band consisting of $N$ levels and a lower band consisting of the other
$M$ levels. We assume that transitions between the upper band and the lower
band are allowed, but those between the energy levels inside the upper band or
the lower band are strictly forbidden.  Such a model may be described by
\begin{equation}\label{MTM}
H_{\rm sys}=\sum_{j=1}^{N+M}\om_j|j\ra\la j|,
\quad L=\sum_{j=1}^M\sum_{k=M+1}^N\ka_{jk}|j\ra\la k|.
\end{equation}
The O-operator is explicitly constructed as a noise-free formation by
$O(t,s)=\sum_{j=1}^M\sum_{k=M+1}^Nf_{jk}(t,s)|j\ra\la k|$ and
$\bar{O}(t)=\sum_{j=1}^M\sum_{k=M+1}^NF_{jk}(t)|j\ra\la k|$ with
$f_{jk}(s,s)=\ka_{jk}$. And we have
\begin{equation}
\pa_tf_{jk}(t,s)=i(\om_k-\om_j)f_{jk}+
\sum_{j'=1}^M\sum_{k'=M+1}^Nf_{jk'}\ka_{j'k'}^*F_{j'k}.
\end{equation}
Therefore, for the cases with the noise-free O-operators, the exact master
equations can be derived directly from Eq.~(\ref{nfme}). It is noted that such
master equations may not be of a standard Lindblad form, but their positivity
is automatically guaranteed by the derivation.

\section{Numerical Results and Discussions}\label{discussion}

\begin{figure}[htbp]
 \centering
 \includegraphics[width=2.7in]{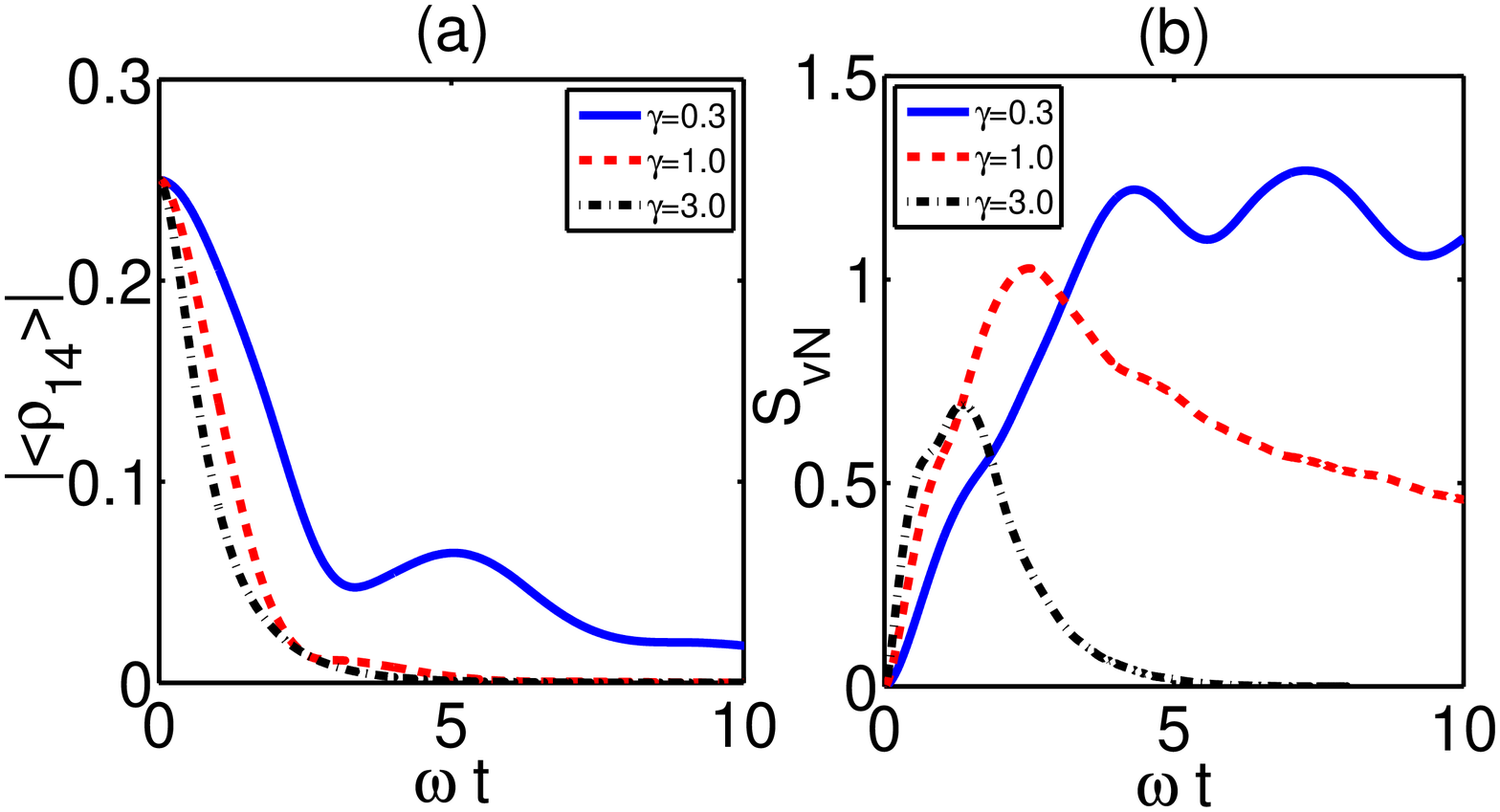}
\caption{(Color online) Coherence and von Neumann entropy \cite{Cohen} of a
dissipative four-level model: $H_{\rm sys}=\om J_z$, and $L=J_-$, and $l=3/2$.
Initially, $\psi_0=(1/2)(|1\ra+|2\ra+|3\ra+|4\ra)$. The time evolution of
coherence $|\la\rho_{14}\ra|$ and entropy $S_{\rm vN}=-\rho\log_2\rho$ is
averaged over $1000$ trajectories for different values of $\gamma$ ($\Ga=\om$)
with the first-order noise term.}
\label{qudit}
\end{figure}

It is known that the stochastic wave-function $\psi_t(z^*)$ in Eq.~(\ref{qsd2})
does not conserve the norm of the wave-function. To efficiently simulate
quantum open systems, one usually uses the nonlinear QSD equation for the
normalized state $\ti{\psi}_t=\frac{\psi_t}{||\psi_t||}$:
\begin{eqnarray}\non
\frac{d}{dt}\ti{\psi}_t&=&\big[-iH_{sys}+(L-\la L\ra_t)\ti{z}^*_t+
\la L^\da\ra_t\big(\bar{O}(t,\ti{z}^*) \\ \non
&-&\la\bar{O}(t,\ti{z}^*)\ra_t\big)-
\left(L^\da\bar{O}(t,\ti{z}^*)-\la L^\da\bar{O}(t,\ti{z}^*)\ra_t\right)
\big]\ti{\psi}_t, \\ \label{nLQSD}
\end{eqnarray}
where $\la L\ra_t=\la\ti{\psi}_t|L^\da|\ti{\psi}_t\ra$ and $\ti{z}^*_t
=z_t^*+\int_0^t\alpha^*(t,s)\la L^\da\ra_sds$ is the shift complex Gaussian
process.

The non-Markovian QSD approach is valid for an arbitrary correlation function.
For simplicity and considering Markov limit, the non-Markovian effect is
modeled by the Lorentz spectral density:
$S(\om)=\frac{1}{2\pi}\frac{\Ga\ga^2}{\ga^2+\om^2}$. Then the correlation
function obtained from the Fourier transformation is given by
$\alpha(t,s)=\int_0^\infty
d\om S(\om)e^{-i\om(t-s)}=\frac{\Ga\ga}{2}e^{-\ga|t-s|}$, where $1/\ga$ is an
important non-Markovian parameter representing the memory time of the
environment. When $\ga\rightarrow\infty$, the correlation function approaches
the Markov limit with $\alpha(t,s)\rightarrow\Ga\delta(t-s)$.

As our first example, we consider the numerical simulation of the dissipative
dynamics of a four-level system with the O-operator given by Eq.~(\ref{O4l}).
In the case of dissipative bath at the zero temperature, the quantum coherence
of the four-level system will decay and the populations of the excited levels
will be transferred to the ground state with time. When the bath is in a
non-Markovian regime with $\gamma=0.3$, it is shown that the decoherence
dynamics deviates from the exponential decay of the Markov case (See the bump
of $\rho_{14}$ in Fig.~\ref{qudit}(a)). Consequently, the decoherence is
delayed as the entanglement between system and bath builds up. We also plot the
von Neumann entropy [$S_{\rm vN}=-{\rm Tr}(\rho\ln \rho)$] (setting the
Boltzmann constant $k_B=1$) in Fig.~\ref{qudit}(b). When the parameter $\ga$
increases to $1.0$ (moderate non-Markovian regime) and $3.0$ (near-Markov
regime), the coherence quickly decays to zero at the time point ($\om t \simeq
5$). However, we see that their entropy curves are different. For example, when
$\ga=3.0$, $S_{\rm vN}$ approaches zero at $\om t \simeq 5$, which means that
the decoherence time between the ground state and the highest energy level and
purification time (That is, $S_{\rm vN} \simeq 0$) coincide. Interestingly, we
can see the $S_{\rm vN}$ is significantly modified by the non-Markovian effect
with $\ga\leqslant1$.

\begin{figure}[htbp]
 \centering
 \includegraphics[width=2.7in]{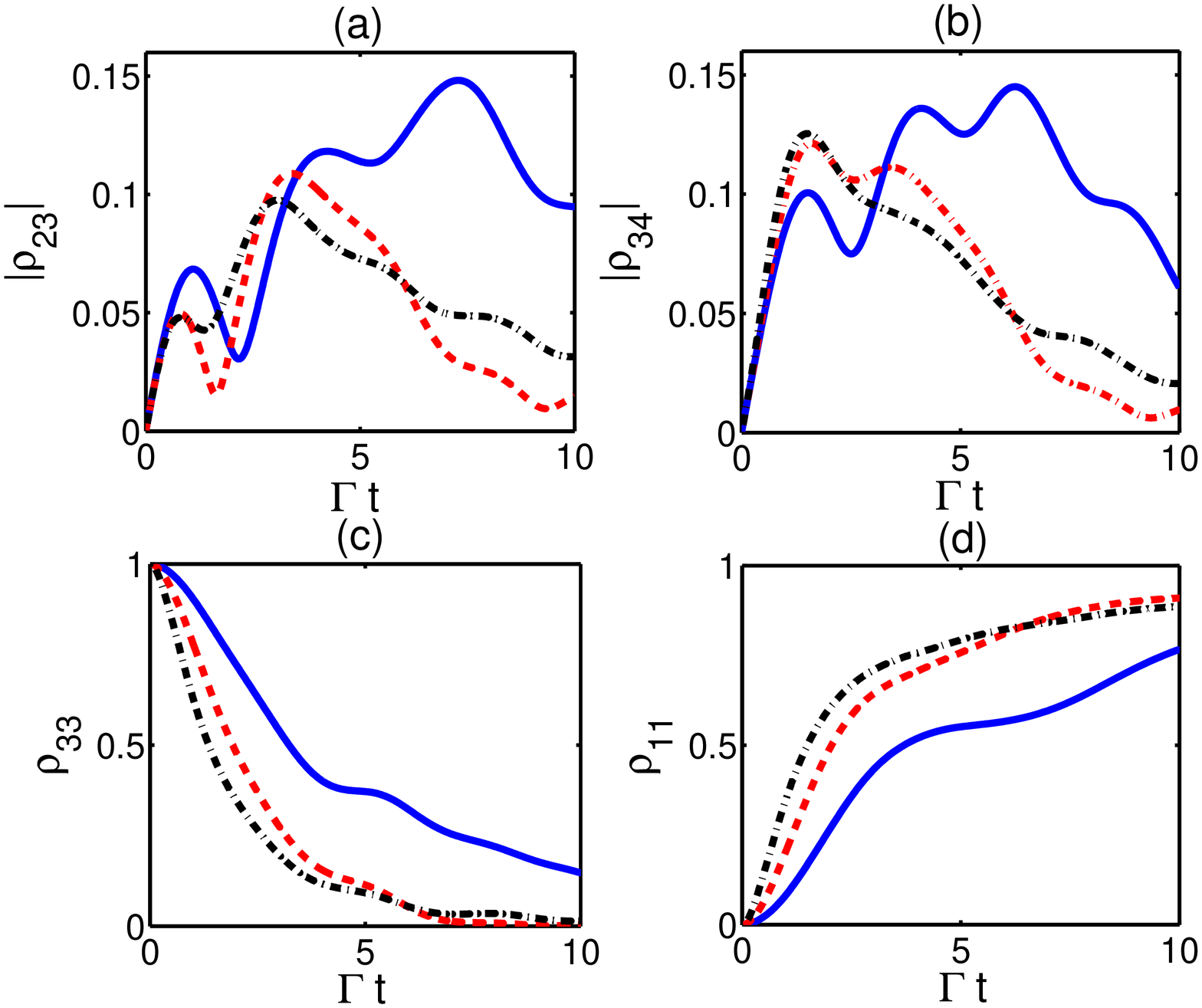}
\caption{(Color online) Coherence and population dynamics of a dissipative
four-level model driven by two external fields. The model is described by
Fig.~\ref{Diag} and Eq.~(\ref{FDA}). The parameters are chosen as follows:
$\om_1=0.1\Ga$, $\om_2=0.3\Ga$, $\om_3=0.6\Ga$, $\om_4=0.2\Ga$,
$\De_2(t)=0.1\Ga e^{2it}T_{23}+h.c.$, $\De_4(t)=0.1\Ga e^{2it}T_{34}+h.c.$,
$\ka_2=0.4$, $\ka_3=0.8$ and $\ka_4=0.3$. Initially, we choose $\psi_0=|3\ra$.
The time evolution of the reduced density matrix is obtained by averaging over
$1000$ trajectories for each $\gamma$ obtained from the QSD equation with the
first-order noise term: (i) $\ga=0.3$, the blue solid lines; (ii) $\ga=1.0$,
the red dashed lines; (iii) $\ga=3.0$, the black dot-dashed
lines.}\label{Fourdri}
\end{figure}

In Fig.~\ref{Fourdri}, we show the dynamics of a driven four-level atomic
system coupled to a dissipative environment with different $\ga$'s. When $t=0$,
the atom is totally populated at the highest level $|3\ra$. With the modulation
of the driving fields $\De_2(t)$ and $\De_4(t)$, Level $2$ and Level $4$ will
be coupled to Level $3$, which helps to establish the coherence terms
$\rho_{23}$ and $\rho_{34}$. Those coherence terms will decay due to the
dissipative channels to Level $1$. From Fig.~\ref{Fourdri}(a) and
Fig.~\ref{Fourdri}(b), when $\ga=0.3$ the non-Markovian environment with long
memory time clearly increases the magnitude of the coherence terms. It should
be noted that the differences between $\ga=1.0$ and $\ga=3.0$ are not
significant since the system is relatively close to the Markov regimes. We can
also see from Fig.~\ref{Fourdri}(c) and Fig.~\ref{Fourdri}(d), that when
$\ga=0.3$, the non-Markovian effect causes population fluctuation of Level $3$
and Level $1$. Typically, we see that the population transfer rates to the
ground state are increased by the shorter environment memory time.

\section{Conclusion}\label{conclusion}

In summary, we have studied the non-Markovian dynamics of multilevel atomic
systems using the non-Markovian quantum trajectory method. The time-local QSD
equations are obtained by the explicit O-operator construction for several
physically interesting models. As shown in this paper, the time-local
O-operators allow numerical simulations to be implemented efficiently for
multilevel open systems. For the multilevel atoms without driving fields, we
show explicitly how to construct the exact O-operators containing a
finite-order noise terms. For the atomic models with multi-transition channels
including the models with driving external fields, we verify that the
O-operator can take a noise-free form, so the exact QSD equations can be easily
established. In the case that the O-operator contains no noise, the exact
convolutionless master equation can be derived. The results of this paper will
open am avenue to exploring novel and fascinating phenomena in multilevel
atom-field interaction problems where the non-Markovian features are important.

\begin{acknowledgements}
This work has benefited from the interesting discussions with Prof. J. H.
Eberly and Prof. B. L. Hu. We acknowledge grant support from the NSF PHY-
0925174, DOD/AF/AFOSR No. FA9550-12-1-0001, The NBRPC No. 2009CB929300, the
NSFC Nos. 91121015 and 11175110 and the MOE No. B06011.
\end{acknowledgements}

\appendix
\section{Proof for the O-operator in Dissipative High-Spin Models}
\label{Oopdicke}

This appendix provides the details of deriving the O-operator for a multilevel
atom dissipative model in Eq.~(\ref{Dick}). The techniques with some necessary
yet not complicated modifications can be used to derive the O-operators for a
general multilevel models including the models with the external driving
fields. For the high-spin model, the system Hamiltonian and the Lindblad
operator may be generally rewritten as
\begin{eqnarray} \non
H_{\rm sys}&=&\sum_{m=1}^{2l+1}C_mH_{\rm sys}^{(m)}, \quad
H_{\rm sys}^{(m)}=|m\ra\la m|, \\ \label{lspin}
L&=&\sum_{n=2}^{2l+1}G_nL_n, \quad L_n=|n-1\ra\la n|.
\end{eqnarray}
When $C_m$'s and $G_n$'s are chosen as the Clebsch-Gordan coefficients, these
operators are referred to the angular momentum operators of spin-$l$. In
practice, Eq.~(\ref{lspin}) can be generalized to a general dissipative model
for a multilevel atom with an arbitrary energy distribution. The model can also
relaxed to the case containing time-dependent damping coefficients.

In what follows, we will show that there are $(2l-k)$ terms containing
$k$th-order noise ($k=0,\cdots, 2l-1$) in the O-operator. Explicitly,
$O(t,s,z^*)$ may be expanded as
\begin{eqnarray}\non
O&=&\sum_{j=1}^{2l}f_j(t,s)O_j^{(0)}+\sum_{j=1}^{2l-1}\int_0^t
p^{(1)}_j(t,s,s_1)z^*_{s_1}ds_1O_j^{(1)} \\ \non
&+&\cdots+\int_0^t\cdots\int_0^tp_1^{(2l-1)}(t,s,s_1,\cdots,
s_{2l-1})z^*_{s_1}\cdots \\ \label{HLA} && z^*_{s_{2l-1}}
ds_1ds_2\cdots ds_{2l-1}O_1^{(2l-1)},
\end{eqnarray}
where the coefficients of $O_j^{(k)}$ ($1\leqslant j\leqslant 2l-k$) are
symmetric functions of $s_1,...s_k$. Subsequently,
\begin{eqnarray}\non
\bar{O}&=&\sum_{j=1}^{2l}F_j(t)O_j^{(0)}+\sum_{j=1}^{2l-1}\int_0^t
P^{(1)}_j(t,s_1)z^*_{s_1}ds_1O_j^{(1)} \\ \non
&+&\cdots+\int_0^t\cdots\int_0^tP_1^{(2l-1)}(t,s_1,\cdots,
s_{2l-1})z^*_{s_1}\cdots \\  && z^*_{s_{2l-1}}
ds_1ds_2\cdots ds_{2l-1}O_1^{(2l-1)}.
\end{eqnarray}
In accord with definition, $F_j(t)\equiv\int_0^tds\alpha(t,s)f_j(t,s)$
and $P_j^{(k)}(t,s_1,\cdots,s_k)\equiv\int_0^tds\alpha(t,s)p_j^{(k)}
(t,s,s_1,\cdots,s_k)$, $k=1,\cdots, 2l-1$.

Clearly, the operators $O_j^{(k)}$ form a set of basis operators:
\begin{equation}\label{Ooplspin}
O_j^{(k)}=|j\ra\la j+k+1|.
\end{equation}
Hence the total number of the basis operators $O_j^{(k)}$ is $(2l+1)l$. In
fact, any linear combinations of operators in Eq.~(\ref{Ooplspin}) may be used
to construct the O-operator as long as they are linear-independent and satisfy
Eq.~(\ref{CC}). Here we use Eq.~(\ref{Ooplspin}) for the simplicity in the
following proof. More often the notation $O_j^{(k)}=J_z^{j-1}J_-^{k+1}$
has a more transparent meaning for the angular momentum model.

By substituting Eq.~(\ref{lspin}) into Eq.~(\ref{CC}), the left hand side of
Eq.~(\ref{CC}) becomes
\begin{eqnarray}\non
&&\sum_{j=1}^{2l}\frac{\partial}{\partial t}
f_j(t,s)O_j^{(0)}+\sum_{j=1}^{2l-1}\bigg[\int_0^t
\frac{\partial}{\partial t}p^{(1)}_j(t,s,s_1)z_{s_1}^*ds_1 \\ \non &+&
z_t^*p^{(1)}_j(t,s,t)\bigg]O_j^{(1)} + \cdots
+\bigg[\int_0^t\cdots\int_0^t\frac{\partial}{\partial t} \\ \non &&
p_1^{(2l-1)}(t,s,s_1,\cdots,s_{2l-1})
z_{s_1}^*\cdots z_{s_{2l-1}}^*ds_1\cdots ds_{2l-1}\\ \non &+&(2l-1)
z_t^*\int_0^t\cdots\int_0^tp_1^{(2l-1)}(t,s,s_1,\cdots,s_{2l-2},t)
\\ \label{parO} &\times&
z_{s_1}^*\cdots z_{s_{2l-2}}^*ds_1\cdots ds_{2l-2}\bigg]O_1^{(2l-1)}.
\end{eqnarray}
The right hand side of the equation consists of four terms. It can be easily
seen that each of them can be expressed as a linear combination of the
operators given in Eq.~(\ref{Ooplspin}). The construction of the O-operator can
be summarized in the following four crucial observations:

(i) $[-iH_{\rm sys},O(t,s,z^*)]$ consisting of $[H_{\rm
sys}^{(m)},O_j^{(k)}]$ can be decomposed as
\begin{equation}
|j\ra\la j+k+1|(\de_{j,m}-\de_{j+k+1,m})=O_j^{(k)}(\de_{j,m}-\de_{j+k+1,m}).
\end{equation}

(ii) $[Lz_t^*,O(t,s,z^*)]$ consisting of the noise term $z_t^*[L_n,O_j^{(k)}]$
turns out to be
\begin{eqnarray}\non
&&z_t^*(|j-1\ra\la j+k+1|\de_{j,n}-|j\ra\la j+k+2|\de_{j+k+1,n-1}) \\
&=&z_t^*(O_{j-1}^{(k+1)}\de_{j,n}-O_j^{(k+1)}\de_{j+k,n-2}).
\end{eqnarray}
These terms correspond to those with $z_t^*$ in Eq.~(\ref{parO}), which will
appear in the boundary conditions between $f_j$'s and $p_j^{(k)}$'s. When $n$
in $L_n$ runs from $2$ to $2l+1$, $O_j^{(k+1)}$'s will present themselves
successively with $j$ in $O_j^{(k)}$ running from $1$ to $2l-k$. It is also a
necessary requirement for the initial condition of O-operator:
$O(s,s,z^*)=L=\sum_{n=2}^{2l+1}G_nL_n$.

(iii) $[-L^\da\bar{O},O]$ is consisted by $[O_j^{(k)},|n\ra\la
n-1|O_{j'}^{(k')}]$, which turns out to be
\begin{equation}
O_j^{(k+k')}\delta_{j+k,j'}-O_{j'+1}^{(k+k')}\delta_{j'+k'+1,j}.
\end{equation}

(iv) For $-L^\da\frac{\de\bar{O}(t,z^*)}{\de z_s^*}$, we only need to consider
the terms with $k\geqslant 1$. The differential functional leaves the operator
unchanged. Then one typical component is
\begin{equation}
-L_n^\da O_j^{(k)}=-|j+1\ra\la j+k+1|\de_{n,j+1}=-O_{j+1}^{(k-1)}.
\end{equation}
It contributes to the presence of $O_j^{(0)}$'s.

To summarize the last four steps, we have proved the existence and
constitutions of the O-operator for the angular momentum model of the
multi-level atom. By introducing Eq.~(\ref{Ooplspin}), the consistency
condition equation (\ref{CC}) is shown to be consistent and complete.

Now we consider a general four-level atom dissipative model with $l=3/2$ in
Eq.~(\ref{lspin}). The group of closed differential equations for the
coefficients can be derived through the above analysis. They are given by,
\begin{eqnarray}\non
\pa_tf_1&=&-i(C_1-C_2)f_1+G_2^*f_1F_1, \\ \non
\pa_tf_2&=&-i(C_2-C_3)f_2+G_3^*f_2F_2-G_2^*f_2F_1-G_2^*P_1^{(1)}, \\ \non
\pa_tf_3&=&-i(C_3-C_4)f_3+G_4^*f_3F_3-G_3^*f_3F_2-G_3^*P_2^{(1)}, \\ \non
\pa_tp_1^{(1)}&=&-i(C_1-C_3)p_1^{(1)}+G_2^*f_1P_1^{(1)}
+G_3^*F_2p_1^{(1)}, \\ \non
\pa_tp_2^{(1)}&=&-i(C_2-C_4)p_2^{(1)}+G_3^*f_2P_2^{(1)}+G_4^*F_3p_2^{(1)}
\\ \non &-&G_2^*F_1p_2^{(1)}-G_2^*f_3P_1^{(1)}-2G_2^*P_1^{(2)}, \\ \non
\pa_tp_1^{(2)}&=&-i(C_1-C_4)p_1^{(2)}+G_2^*f_1P_1^{(2)}+G_3^*P_2^{(1)}p_1^{(1)}
\\ &+&G_4^*F_3p_1^{(2)}.
\end{eqnarray}
Here the boundary conditions are,
\begin{eqnarray}\non
2p_1^{(2)}(t,s,s_1,t)&=&G_2p_2^{(1)}(t,s,s_1)-G_4p_1^{(1)}(t,s,s_1), \\ \non
p_1^{(1)}(t,s,t)&=&G_2f_2(t,s)-G_3f_1(t,s), \\
p_2^{(1)}(t,s,t)&=&G_3f_3(t,s)-G_4f_2(t,s).
\end{eqnarray}

The spin-$3/2$ dissipative model [See the O-operator in Eq.~(\ref{O4l})] can
be solved by setting $C_m=-l-1+m$, $m=1,2,3,4$ and $G_2=G_4=\sqrt{3}$,
$G_3=2$.

\end{document}